\documentclass[aps,prl,twocolumn,superscriptaddress,showpacs,preprintnumbers,amsmath,amssymb]{revtex4}

\usepackage{graphicx} % Include figure files
\usepackage{dcolumn}  % Align table columns on decimal point
\usepackage{gensymb}

\graphicspath{{ps}}

\begin{document}

\preprint{\vbox{ \hbox{   }
	          	\hbox{Belle Preprint 2018-08}
                         \hbox{KEK Preprint 2018-2}
}}

\title{ \quad\\[1.0cm] Observation of \boldmath$\Upsilon(4S)\to\eta'\Upsilon(1S)$}

\noaffiliation
\affiliation{University of the Basque Country UPV/EHU, 48080 Bilbao}
\affiliation{Beihang University, Beijing 100191}
\affiliation{Budker Institute of Nuclear Physics SB RAS, Novosibirsk 630090}
\affiliation{Faculty of Mathematics and Physics, Charles University, 121 16 Prague}
\affiliation{University of Cincinnati, Cincinnati, Ohio 45221}
\affiliation{Deutsches Elektronen--Synchrotron, 22607 Hamburg}
\affiliation{University of Florida, Gainesville, Florida 32611}
\affiliation{Fudan University, Shanghai 200443}
\affiliation{Gifu University, Gifu 501-1193}
\affiliation{SOKENDAI (The Graduate University for Advanced Studies), Hayama 240-0193}
\affiliation{Gyeongsang National University, Chinju 660-701}
\affiliation{Hanyang University, Seoul 133-791}
\affiliation{University of Hawaii, Honolulu, Hawaii 96822}
\affiliation{High Energy Accelerator Research Organization (KEK), Tsukuba 305-0801}
\affiliation{J-PARC Branch, KEK Theory Center, High Energy Accelerator Research Organization (KEK), Tsukuba 305-0801}
\affiliation{IKERBASQUE, Basque Foundation for Science, 48013 Bilbao}
\affiliation{Indian Institute of Science Education and Research Mohali, SAS Nagar, 140306}
\affiliation{Indian Institute of Technology Guwahati, Assam 781039}
\affiliation{Indian Institute of Technology Hyderabad, Telangana 502285}
\affiliation{Indian Institute of Technology Madras, Chennai 600036}
\affiliation{Indiana University, Bloomington, Indiana 47408}
\affiliation{Institute of High Energy Physics, Chinese Academy of Sciences, Beijing 100049}
\affiliation{Institute of High Energy Physics, Vienna 1050}
\affiliation{Institute for High Energy Physics, Protvino 142281}
\affiliation{INFN - Sezione di Napoli, 80126 Napoli}
\affiliation{INFN - Sezione di Torino, 10125 Torino}
\affiliation{Advanced Science Research Center, Japan Atomic Energy Agency, Naka 319-1195}
\affiliation{J. Stefan Institute, 1000 Ljubljana}
\affiliation{Kanagawa University, Yokohama 221-8686}
\affiliation{Institut f\"ur Experimentelle Teilchenphysik, Karlsruher Institut f\"ur Technologie, 76131 Karlsruhe}
\affiliation{Kennesaw State University, Kennesaw, Georgia 30144}
\affiliation{King Abdulaziz City for Science and Technology, Riyadh 11442}
\affiliation{Department of Physics, Faculty of Science, King Abdulaziz University, Jeddah 21589}
\affiliation{Korea Institute of Science and Technology Information, Daejeon 305-806}
\affiliation{Korea University, Seoul 136-713}
\affiliation{Kyoto University, Kyoto 606-8502}
\affiliation{Kyungpook National University, Daegu 702-701}
\affiliation{\'Ecole Polytechnique F\'ed\'erale de Lausanne (EPFL), Lausanne 1015}
\affiliation{P.N. Lebedev Physical Institute of the Russian Academy of Sciences, Moscow 119991}
\affiliation{Faculty of Mathematics and Physics, University of Ljubljana, 1000 Ljubljana}
\affiliation{Ludwig Maximilians University, 80539 Munich}
\affiliation{Luther College, Decorah, Iowa 52101}
\affiliation{University of Malaya, 50603 Kuala Lumpur}
\affiliation{University of Maribor, 2000 Maribor}
\affiliation{Max-Planck-Institut f\"ur Physik, 80805 M\"unchen}
\affiliation{School of Physics, University of Melbourne, Victoria 3010}
\affiliation{University of Mississippi, University, Mississippi 38677}
\affiliation{University of Miyazaki, Miyazaki 889-2192}
\affiliation{Moscow Physical Engineering Institute, Moscow 115409}
\affiliation{Moscow Institute of Physics and Technology, Moscow Region 141700}
\affiliation{Graduate School of Science, Nagoya University, Nagoya 464-8602}
\affiliation{Nara Women's University, Nara 630-8506}
\affiliation{National Central University, Chung-li 32054}
\affiliation{National United University, Miao Li 36003}
\affiliation{Department of Physics, National Taiwan University, Taipei 10617}
\affiliation{H. Niewodniczanski Institute of Nuclear Physics, Krakow 31-342}
\affiliation{Nippon Dental University, Niigata 951-8580}
\affiliation{Niigata University, Niigata 950-2181}
\affiliation{Novosibirsk State University, Novosibirsk 630090}
\affiliation{Osaka City University, Osaka 558-8585}
\affiliation{Pacific Northwest National Laboratory, Richland, Washington 99352}
\affiliation{Panjab University, Chandigarh 160014}
\affiliation{Peking University, Beijing 100871}
\affiliation{Punjab Agricultural University, Ludhiana 141004}
\affiliation{Theoretical Research Division, Nishina Center, RIKEN, Saitama 351-0198}
\affiliation{University of Science and Technology of China, Hefei 230026}
\affiliation{Showa Pharmaceutical University, Tokyo 194-8543}
\affiliation{Soongsil University, Seoul 156-743}
\affiliation{University of South Carolina, Columbia, South Carolina 29208}
\affiliation{Stefan Meyer Institute for Subatomic Physics, Vienna 1090}
\affiliation{Sungkyunkwan University, Suwon 440-746}
\affiliation{School of Physics, University of Sydney, New South Wales 2006}
\affiliation{Department of Physics, Faculty of Science, University of Tabuk, Tabuk 71451}
\affiliation{Tata Institute of Fundamental Research, Mumbai 400005}
\affiliation{Department of Physics, Technische Universit\"at M\"unchen, 85748 Garching}
\affiliation{Toho University, Funabashi 274-8510}
\affiliation{Department of Physics, Tohoku University, Sendai 980-8578}
\affiliation{Earthquake Research Institute, University of Tokyo, Tokyo 113-0032}
\affiliation{Department of Physics, University of Tokyo, Tokyo 113-0033}
\affiliation{Tokyo Institute of Technology, Tokyo 152-8550}
\affiliation{Tokyo Metropolitan University, Tokyo 192-0397}
\affiliation{University of Torino, 10124 Torino}
\affiliation{Virginia Polytechnic Institute and State University, Blacksburg, Virginia 24061}
\affiliation{Wayne State University, Detroit, Michigan 48202}
\affiliation{Yonsei University, Seoul 120-749}
  \author{E.~Guido}\affiliation{INFN - Sezione di Torino, 10125 Torino} % Torino
  \author{R.~Mussa}\affiliation{INFN - Sezione di Torino, 10125 Torino} % Torino
  \author{U.~Tamponi}\affiliation{INFN - Sezione di Torino, 10125 Torino}\affiliation{University of Torino, 10124 Torino} % Torino
  \author{H.~Aihara}\affiliation{Department of Physics, University of Tokyo, Tokyo 113-0033} % Tokyo
  \author{S.~Al~Said}\affiliation{Department of Physics, Faculty of Science, University of Tabuk, Tabuk 71451}\affiliation{Department of Physics, Faculty of Science, King Abdulaziz University, Jeddah 21589} % Tabuk
  \author{D.~M.~Asner}\affiliation{Pacific Northwest National Laboratory, Richland, Washington 99352} % PNNL
  \author{H.~Atmacan}\affiliation{University of South Carolina, Columbia, South Carolina 29208} % SouthCarolina
  \author{V.~Aulchenko}\affiliation{Budker Institute of Nuclear Physics SB RAS, Novosibirsk 630090}\affiliation{Novosibirsk State University, Novosibirsk 630090} % BINP
  \author{T.~Aushev}\affiliation{Moscow Institute of Physics and Technology, Moscow Region 141700} % MIPT
  \author{R.~Ayad}\affiliation{Department of Physics, Faculty of Science, University of Tabuk, Tabuk 71451} % Tabuk
% ...
% BINP
\collaboration{The Belle Collaboration}

\begin{abstract}
We report the first observation of the hadronic transition $\Upsilon(4S)\to\eta'\Upsilon(1S)$, using 496  fb$^{-1}$ data collected at the $\Upsilon(4S)$ resonance with the Belle detector at the KEKB asymmetric-energy $e^{+}e^{-}$ collider. We reconstruct the $\eta'$ meson through its decays to $\rho^0\gamma$ and to $\pi^+\pi^-\eta$, with $\eta\to\gamma\gamma$. We measure: ${\cal B}(\Upsilon(4S)\to\eta'\Upsilon(1S))=(3.43\pm 0.88 {\rm(stat.)} \pm 0.21 {\rm(syst.)})\times10^{-5}$, with a significance of 5.7$\sigma$.
\end{abstract}

\pacs{14.40.Pq, 13.25.Gv}

\maketitle

\tighten

{\renewcommand{\thefootnote}{\fnsymbol{footnote}}}
\setcounter{footnote}{0}

One of the major challenges in particle physics is the treatment of non-perturbative QCD~\cite{ref:Brambilla2014}.
Quarkonia, thanks to their intrinsic multi-scale behavior, are one of the most promising and clean laboratories in which to explore these dynamics~\cite{ref:Brambilla2010}.
In particular, hadronic transitions between bottomonia have been, in the past few years, a fertile field for both experiment and theory. On the basis of heavy quark spin symmetry, the QCD multipole expansion (QCDME) model predicts that $\eta$ transitions should be suppressed relative to dipion transitions~\cite{ref:QCDME}. Several recent results~\cite{ref:BaBar4S,ref:Belle4S,ref:Belle4Setahb,ref:Belle5Seta1D} challenge this long-standing expectation. Following these measurements, it has been argued that the light-quark degrees of freedom actively intervene in the transitions~\cite{ref:Voloshin}.

Few processes for the $\Upsilon(4S)$ decaying to the non-$B\bar B$ system have been measured thus far~\cite{ref:PDG2017}.
There have been no searches for the kinematically allowed transition $\Upsilon(4S)\to\eta'\Upsilon(1S)$ , which is expected to be enhanced just as $\Upsilon(4S)\to\eta\Upsilon(1S)$~\cite{ref:Voloshin}, where the relative strength of the $\eta'$ and $\eta$ transitions depends on the relative $u\bar u + d\bar d$ content of the mesons, and is predicted to range between 20 and 60$\%$. In contrast, a significant dominance of the $\eta'$ transition is predicted by QCDME models.
In the charmonium sector, searches for $\psi(4160)\to\eta' J/\psi$ and $Y(4260)\to\eta' J/\psi$ transitions have been made by CLEO~\cite{ref:CLEOcharm} without observation of significant signals, while the observation of $e^+e^-\to\eta' J/\psi$ at center-of-mass energy of 4.226 GeV and 4.258 GeV has been reported by BESIII~\cite{ref:BES3charm}.

In this Letter, we present the first observation of the transition $\Upsilon(4S)\to\eta'\Upsilon(1S)$. The $\Upsilon(1S)$ meson is reconstructed via its leptonic decay to two muons, which is considerably cleaner than the di-electron mode. The $\eta'$ meson is reconstructed via its decays to $\rho^0\gamma$ and to $\pi^+\pi^-\eta$, with the $\eta$ meson reconstructed as two photons. 

We use a sample of $(538\pm8)\times10^6$ $\Upsilon(4S)$ mesons, corresponding to an integrated luminosity of 496 fb$^{-1}$, collected by the Belle experiment at the KEKB asymmetric-energy $e^+e^-$ collider~\cite{ref:KEKB,ref:KEKB_bis}. 
In addition, a data sample corresponding to 56 fb$^{-1}$, collected about 60~MeV below the resonance, is used to estimate the background contribution. 

The Belle detector (described in detail elsewhere~\cite{ref:Belle,ref:Belle_bis})  is a large-solid-angle magnetic spectrometer that consists of a silicon vertex detector, a 50-layer central drift chamber (CDC), an array of aerogel threshold Cherenkov counters (ACC), a barrel-like arrangement of time-of-flight scintillation counters, and an electromagnetic calorimeter comprised of CsI(Tl) crystals (ECL) located inside a super-conducting solenoid coil that provides a 1.5~T magnetic field.  An iron flux-return located outside of the coil (KLM) is instrumented to detect $K_L^0$ mesons and to identify muons. 

Monte Carlo (MC) simulated events are used for the efficiency determination and the selection optimization; these are generated using \texttt{EvtGen}~\cite{ref:EvtGen} and simulated to model the detector response using \texttt{GEANT}3~\cite{ref:GEANT}. The changing detector performance and accelerator conditions are taken into account in the simulation.
The distributions of generated dimuon decays incorporate the $\Upsilon(1S)$ polarization. The angular distribution in the $\Upsilon(4S)\to\eta'\Upsilon(1S)$ transition is simulated as a vector decaying to a pseudoscalar and a vector. The $\eta'\to\pi^+\pi^-\eta$ and the $\eta\to\gamma\gamma$ decays are generated uniformly in phase space, while the $\eta'\to\rho^0\gamma\to\pi^+\pi^-\gamma$ decay is generated assuming the appropriate helicity.
Final state radiation effects are modeled in the generator by \texttt{PHOTOS}~\cite{ref:PHOTOS}.

Charged tracks must originate from a cylindrical region of length $\pm 5\,\rm cm$ along the $z$ axis (opposite the positron beam) and radius 1 cm in the transverse plane, centered on the $e^+e^-$ interaction point, and must have a transverse momentum ($p_{\rm T}$) greater than 0.1 GeV/$c$. Charged particles are assigned a likelihood ${\cal L}_i$, with $i = \mu, \pi, K$~\cite{ref:muID}, based on the range of the particle extrapolated from the CDC through the KLM; particles are identified as muons if the likelihood ratio ${\cal P}_{\mu} = {\cal L}_\mu/({\cal L}_\mu + {\cal L}_\pi + {\cal L}_K)$ exceeds 0.8, corresponding to a muon efficiency of about 91.5$\%$ over the polar angle range $20\degree \leq \theta \leq 155\degree$ and the momentum range 0.7 GeV/$c \leq p \leq 3.0$ GeV/$c$ in the laboratory frame. Electron identification uses a similar likelihood ratio ${\cal P}_e$ based on CDC, ACC, and ECL information~\cite{ref:eID}. Charged particles that are not identified as muons and having a likelihood ratio ${\cal P}_e<0.1$ are treated as pions. 
Calorimeter clusters not associated with reconstructed charged tracks and with energies greater than 50 MeV are classified as photon candidates.
Pairs of oppositely charged tracks, of which at least one is positively identified as a muon, are selected as dimuon candidates.
Pairs of oppositely charged tracks, both classified as pions, are selected as dipion candidates.
Retained events contain one dimuon candidate and one dipion candidate. 
For $\eta'\to\rho^0\gamma$ decays, hereinafter labeled as $2\pi1\gamma$, only events with at least one photon and with the photon-dipion invariant mass within 50 MeV/$c^2$ ($\pm3\sigma$) of the nominal $\eta'$ mass~\cite{ref:PDG2017} are retained. Similarly, for $\eta'\to\pi^+\pi^-\eta$, $\eta\to\gamma\gamma$ decay chain, hereinafter labeled as $2\pi2\gamma$, only events with at least two photons having an invariant mass within 50 MeV/$c^2$ ($\pm3\sigma$) of the nominal $\eta$ mass~\cite{ref:PDG2017}, and with an invariant-mass difference $M(\pi^+\pi^-\gamma\gamma) - M(\gamma\gamma)$ within 20 MeV/$c^2$ ($\pm3\sigma$) of the nominal value are considered. In $2\pi1\gamma$ ($2\pi2\gamma$) final states, 1.2 (1.4) candidates per event are present on average, where the multiplicity is due to the photon(s). The ambiguity is resolved by choosing the one whose reconstructed $\eta'$ mass is closest to the nominal value. This choice has an efficiency of $\sim90\%$ on the MC-simulated signal samples.
The events with $|\sqrt{s}-M(\Upsilon(1S)\eta')c^2|<150$ MeV, where $M(\Upsilon(1S)\eta')=M(\mu^+\mu^-\pi^+\pi^-\gamma)$ [$M(\mu^+\mu^-\pi^+\pi^-\gamma\gamma)$] in the $2\pi1\gamma$ [$2\pi2\gamma$] final state and $\sqrt{s}$ is the center-of-mass (CM) $e^+e^-$ energy, are retained.

The kinematic bound expressed by the quantity $p_{\rm KB} = p(\mu\mu)_{CM} - (s-M(\mu\mu)^2c^4)/(2c\sqrt{s})$, where $p(\mu\mu)_{CM}$ is the CM momentum of the dimuon system, is constrained to negative values for  signal events, and is used to reject part of the background contribution due to QED processes ($e^+e^-\to e^+e^-(\gamma)$ and $e^+e^-\to \mu^+\mu^-(\gamma)$).
Further reductions of QED processes and of cosmic background events are achieved by requiring the opening angle of the charged pion candidates in the CM frame to satisfy $|\cos\theta(\pi\pi)_{\rm CM}|<0.9$.

The $2\pi1\gamma$ final state  has contributions from dipion transitions to the $\Upsilon(1S)$ resonance from either $\Upsilon(2S,3S)$ resonances produced in initial state radiation (ISR) events or the $\Upsilon(4S)$ resonance in which a random photon is incorporated into the $\eta'$ candidate. The high production cross section values~\cite{ref:ISRlumi} and decay rates~\cite{ref:PDG2017} make these processes competitive with the signal transition, and particular care is needed to reduce them to negligible levels. A Boosted Decision Tree (BDT) method, as implemented in the Toolkit for Multivariate Data Analysis package~\cite{ref:TMVA}, is trained to separate the signal events from those due to dipion transitions. The performance of the classifier is optimized and tested using MC-simulated samples for both the signal and dipion transitions.
The input variables used to construct the BDT are the difference between invariant masses $\Delta M_{\pi\pi} = M(\mu^+\mu^-\pi^+\pi^-) - M(\mu^+\mu^-)$ and the total reconstructed mass of the event $M(\mu^+\mu^+\pi^+\pi^-\gamma)$. The highest discrimination is provided by $\Delta M_{\pi\pi}$. This variable is broadly distributed for signal events, and instead assumes the values $563.0\pm0.4$ MeV/$c^{2}$, $894.9\pm0.6$ MeV/$c^{2}$, and $1119.1\pm1.2$ MeV/$c^{2}$, for $\Upsilon(2S)$, $\Upsilon(3S)$, and $\Upsilon(4S)\to\pi^+\pi^-\Upsilon(1S)$, respectively~\cite{ref:PDG2017}, with experimental resolutions of a few MeV/$c^{2}$. It has been verified that, with respect to a cut-based approach, the BDT method enhances the dipion rejection retaining a higher signal efficiency.
The reconstructed invariant mass of the $\eta'$ candidate must lie within 0.93 GeV/$c^{2} < M(\pi^+\pi^-\gamma) < 0.98$ GeV/$c^{2}$, which retains 90$\%$ of signal events.

The overall selection efficiencies for the signal events in the $2\pi1\gamma$ and $2\pi2\gamma$ final states are $\epsilon=(17.64\pm0.05)\%$ and $(5.02\pm0.03)\%$, respectively, as determined from MC-simulated samples. The selection efficiency for  $\Upsilon(2S,3S,4S)\to\pi^+\pi^-\Upsilon(1S)$ events is in the range of $10^{-6}-10^{-4}$, making their contribution negligible. The contributions from these and other background sources are measured with a data sample collected below the $\Upsilon(4S)$ resonance; a fraction of less than ${\sim}10^{-8}$ of the data remains in the $2\pi1\gamma$ final state, while no events are present in the $2\pi2\gamma$ final state.

\begin{figure}[htb]
\includegraphics[width=0.44\textwidth]{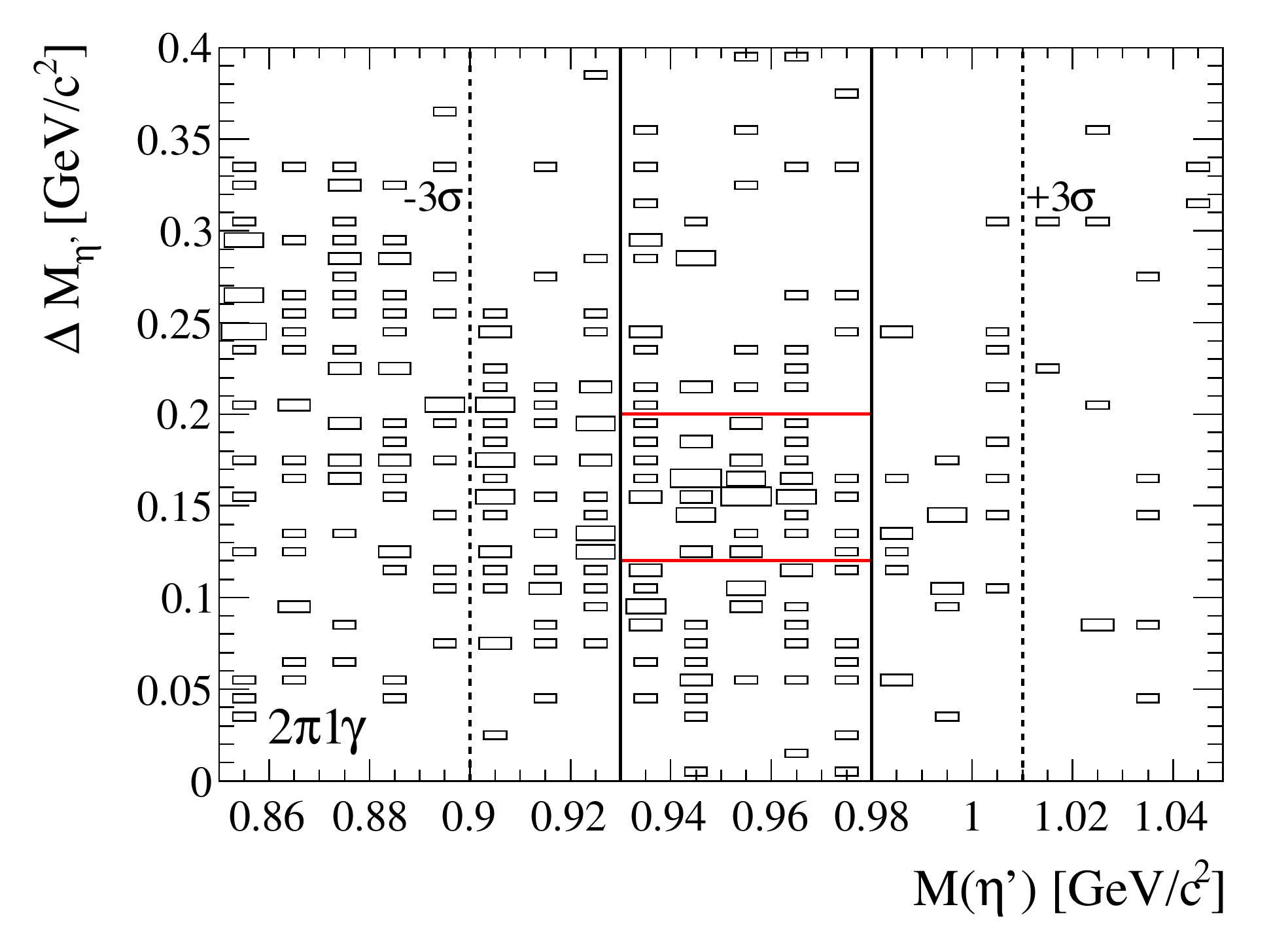}\\
\caption{Distribution of $\Delta M_{\eta'}$ versus $M(\eta')$ for the selected events (binned into the boxes) in the $2\pi1\gamma$ final state. The vertical dashed lines show the $\pm3\sigma$ selected region. The signal-selection region of  0.93 GeV/$c^{2} <M(\pi^+\pi^-\gamma)<0.98$ GeV/$c^{2}$  is bounded by the vertical solid lines. The two-dimensional region where 97\% of the signal events are expected is bounded by these vertical lines and the two red horizontal lines.}\label{fig:fig1}
\end{figure}

\begin{figure}[htb]
\includegraphics[width=0.44\textwidth]{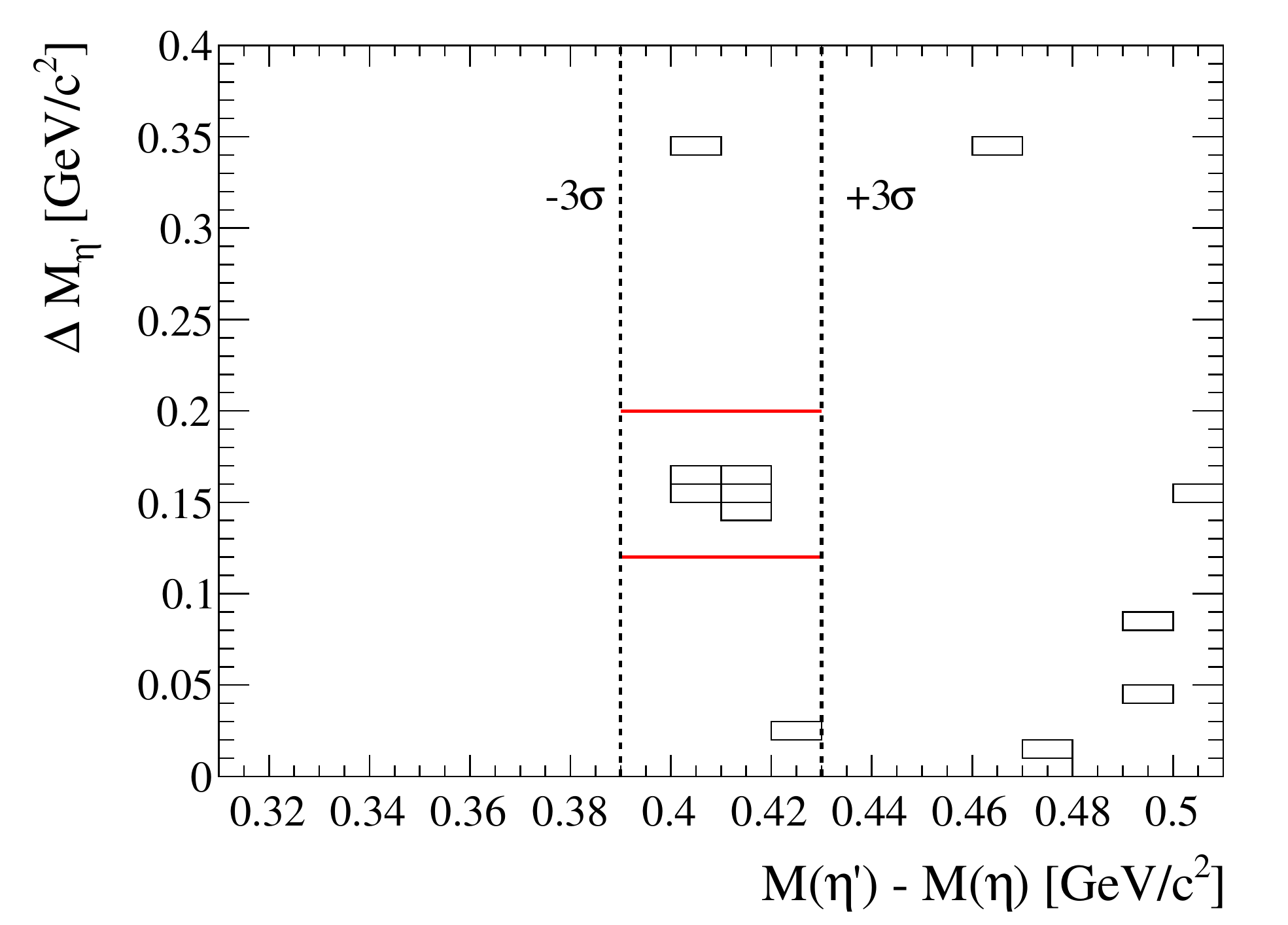}\\
\caption{Distribution of $\Delta M_{\eta'}$ versus $M(\eta') - M(\eta)$ for the selected events (binned into the boxes) in the $2\pi2\gamma$ final state. The vertical dashed lines show the $\pm3\sigma$ selected region. The two-dimensional region where 97\% of the signal events are expected is bounded by these vertical lines and the two red horizontal lines.}\label{fig:fig2}
\end{figure}

The signal events are identified by the variable:
\begin{equation}
\Delta M_{\eta'} = M(\Upsilon(4S)) - M(\Upsilon(1S)) - M(\eta'),
\end{equation}
where $M(\Upsilon(1S))=M(\mu^+\mu^-)$ in both final states; for the $2\pi 1\gamma$ [$2\pi 2\gamma$] final state,  $M(\Upsilon(4S))=M(\mu^+\mu^-\pi^+\pi^-\gamma)~[M(\mu^+\mu^-\pi^+\pi^-\gamma\gamma)]$ and $M(\eta')=M(\pi^+\pi^-\gamma)$ [$M(\pi^+\pi^-\gamma\gamma)$]. The expected resolution for the signal is 7--8 MeV/$c^2$, depending on the reconstructed $\eta'$ decay mode.
The distribution of $\Delta M_{\eta'}$ versus $M(\eta')$ [$M(\eta') - M(\eta)$]  for the $2\pi1\gamma$ [$2\pi2\gamma$] candidates is shown in Fig.~\ref{fig:fig1} [Fig.~\ref{fig:fig2}] in a broad range of the abscissa in order to illustrate the distribution.

\begin{figure}[htb]
\includegraphics[width=0.44\textwidth]{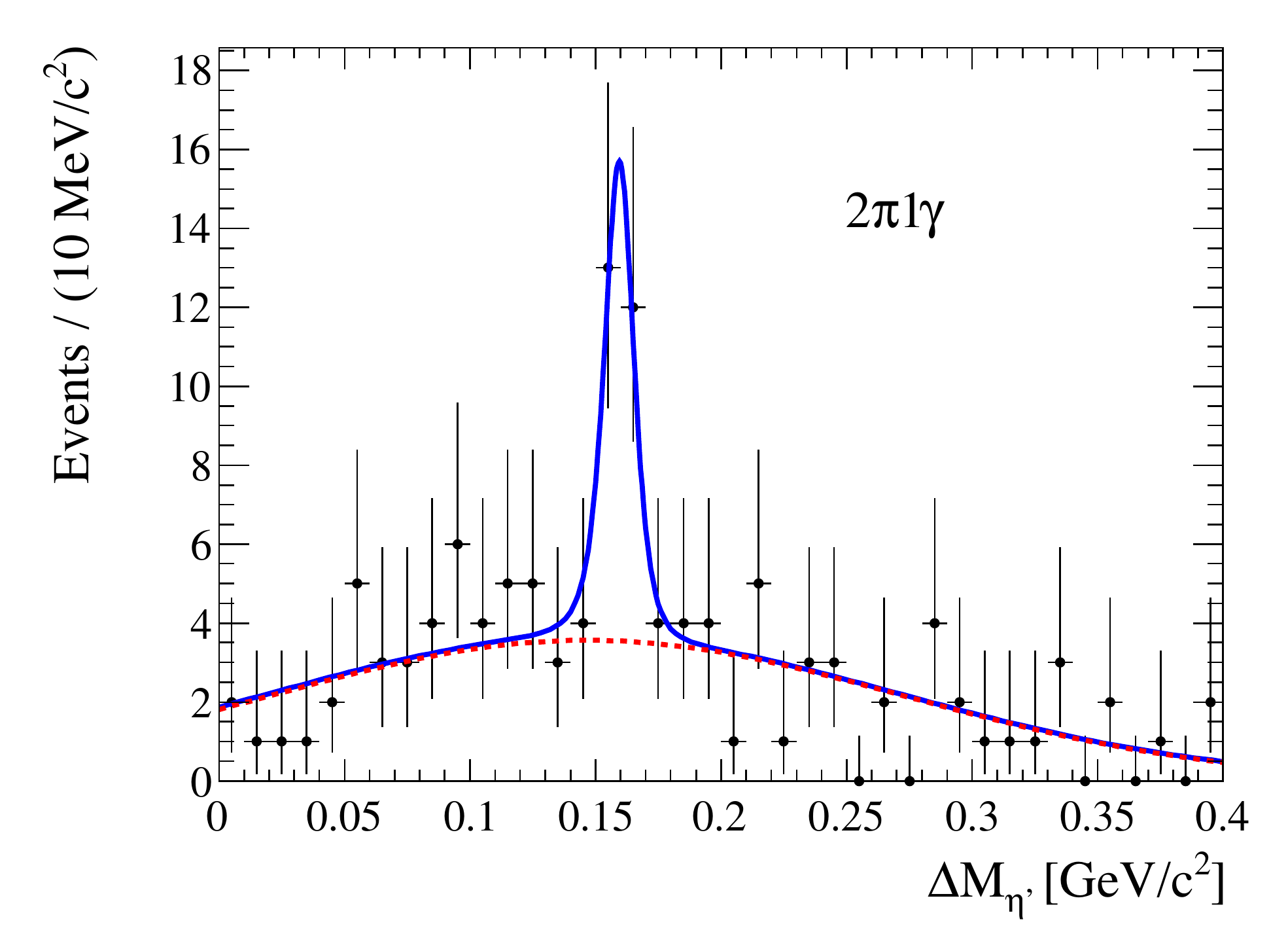}\\
\includegraphics[width=0.44\textwidth]{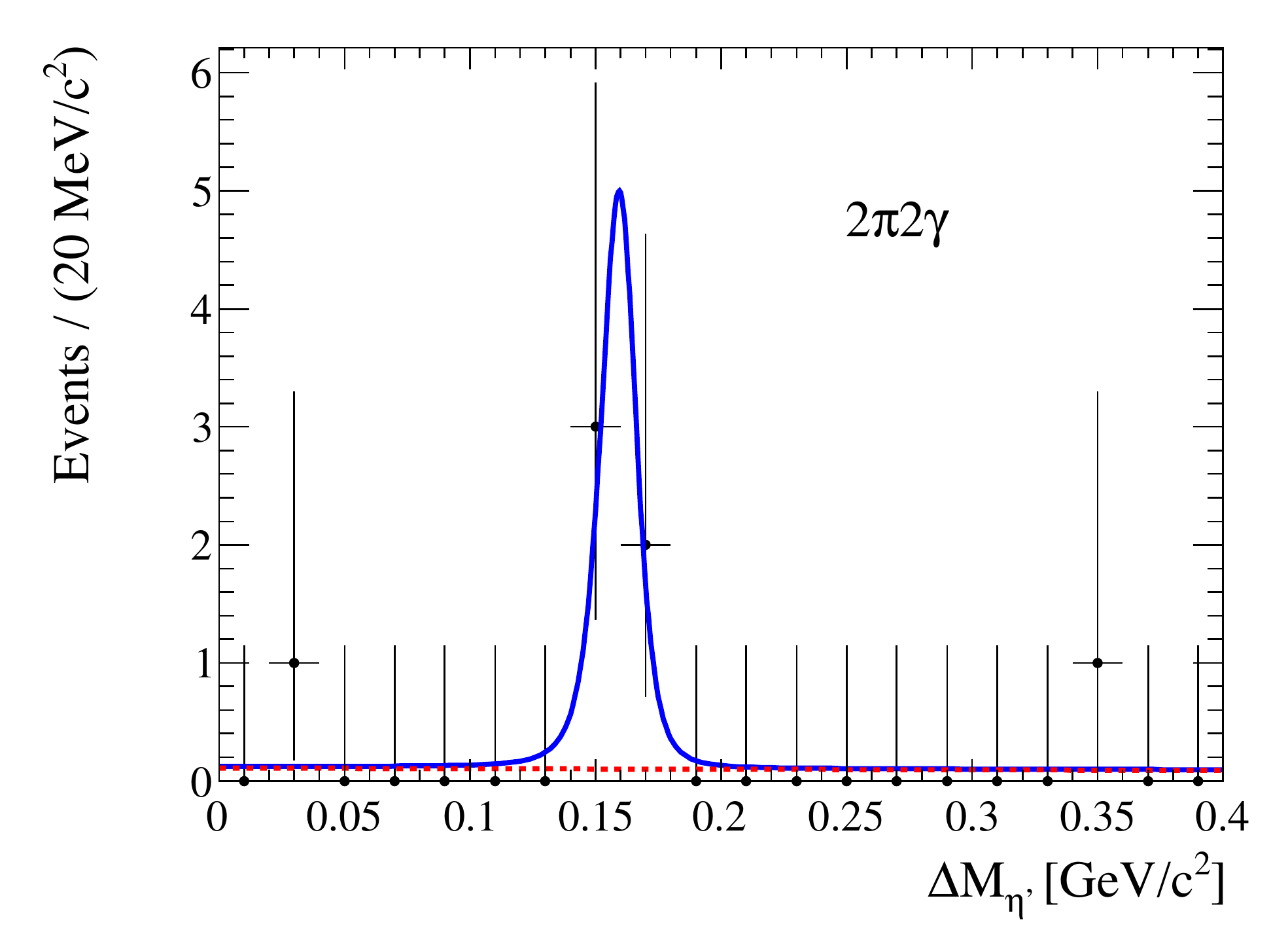}
\caption{Fit to the $\Delta M_{\eta'}$ distribution for $\Upsilon(4S)\to\eta'\Upsilon(1S)$ candidates reconstructed in the $2\pi1\gamma$ (top) and $2\pi2\gamma$ (bottom) final states. Data are shown as points, the solid blue line shows the best fit to the data, while the dashed red line shows the background contribution.}\label{fig:fig3}
\end{figure}

The signal and background yields are determined by an unbinned maximum likelihood fit to the $\Delta M_{\eta'}$ distribution, shown in Fig.~\ref{fig:fig3}.
The signal component is parameterized by a Gaussian-like analytical function
    \begin{equation}
    {\cal F}(x) = \exp\Big\{-\frac{(x-\mu)^2}{2\sigma_{\text{L,R}}^2+\alpha_{\text{L,R}}(x-\mu)^2}\Big\}, \label{eq:cruijffpdf}
    \end{equation}
with mean value $\mu$ and distinct widths, $\sigma_{\text{L,R}}$, and asymmetric-tail parameters, $\alpha_{\rm L,R}$, either side of the peak.
The background is described by a very broad Gaussian (linear) function in the $2\pi1\gamma$ ($2\pi2\gamma$) final state.
The signal shape parameters are fixed to the values determined from the MC-simulated sample. The signal and background yields in the $2\pi1\gamma$ final state are $N_{\rm sig} = 22\pm7$ and $N_{\rm bkg} = 96\pm11$, respectively. In the $2\pi2\gamma$ final state, the signal and background yields are $N_{\rm sig} = 5.0\pm2.3$ and $N_{\rm bkg} = 2.0\pm1.6$, respectively.

The statistical significance of the signal is determined as $\sqrt{2\log[{\cal L}(N_{\rm sig})/{\cal L}(0)]}$, where ${\cal L}(N_{\rm sig})/{\cal L}(0)$ is the ratio between the likelihood values for a fit that includes a signal component versus a fit with only the background hypothesis.
The statistical significance is estimated to be $4.2\sigma$ ($4.1\sigma$) in the $2\pi1\gamma$ ($2\pi2\gamma$) final state.

Several sources of systematic uncertainty affect the branching fraction measurement, including the number of $\Upsilon(4S)$ events, $N_{\Upsilon(4S)}$, ($\pm1.4\%$) and  the values used for the secondary branching fractions, ${\cal B}_{\rm secondary}$ ($\pm2.7\%$ for $2\pi1\gamma$ and $\pm2.6\%$ for $2\pi2\gamma$)~\cite{ref:PDG2017}. The uncertainties in charged track reconstruction ($\pm1.4\%$) and muon identification efficiency ($\pm1.1\%$) are determined by comparing data and MC events using independent control samples. 
The largest contribution to the systematic uncertainty comes from the signal extraction procedure ($\pm6.8\%$ for $2\pi1\gamma$ and $\pm2.0\%$ for $2\pi2\gamma$). The uncertainty due to the choice of signal parameterizations is estimated by changing the functional forms used; the systematic uncertainty for the background form is evaluated by using second-order polynomial or exponential functions, and by varying the range chosen for the fit. An additional uncertainty is related to the chosen values for the signal shape parameters, and is evaluated by repeating the fit while varying each of them by $\pm1\sigma$ with respect to its nominal value. In each case, the uncertainty is estimated as the variation in the signal yield when using an alternate configuration with respect to that obtained with the nominal one.
Not all of the partial width of $\eta'\to\pi^+\pi^-\gamma$ can be explained by a resonant decay through a $\rho^0$~\cite{ref:BES3etap}, but the fractions of the nonresonant and resonant contributions are unmeasured. The potential systematic bias in the signal efficiency due to a non-null fraction of nonresonant decays is estimated by comparing the selection efficiencies between the default resonant sample and a completely nonresonant one. Half of the difference is conservatively assigned as systematic error ($-1.9\%$ for $2\pi1\gamma$).
Other possible sources of systematic uncertainties, due to discrepancies between data and MC in the efficiency of the applied selection requirements or in the photon energy calibration, have been found to be relatively small.
The total systematic uncertainty is obtained by adding in quadrature all of the contributions, and amounts to $7.6\%$ in the $2\pi1\gamma$ final state and $3.5\%$ in the $2\pi2\gamma$ final state.

The value of the branching fraction ${\cal B}$ is calculated as:
\begin{equation}
{\cal B} =\frac{N_{\rm sig}}{\epsilon \times N_{\Upsilon(4S)} \times {\cal B}_{\rm secondary}}.
\end{equation}
We measure ${\cal B} = (3.19\pm0.96({\rm stat.})\pm0.24({\rm syst.}))\times10^{-5}$ in the $2\pi1\gamma$ final state, and ${\cal B} = (4.53\pm2.12({\rm stat.})\pm0.16({\rm syst.}))\times10^{-5}$ in the $2\pi2\gamma$ final state.
The measurements obtained from the two independent subsamples are combined in a weighted average, where the weight is the inverse of the squared sum of the statistical and systematic uncertainties on each yield, considering only the systematic contributions that are uncorrelated between the two channels. The systematic uncertainties in common between the two channels are then added in quadrature to obtain the total uncertainty.
The measured branching fraction is: ${\cal B}(\Upsilon(4S)\to\eta'\Upsilon(1S)) = (3.43\pm 0.88{\rm (stat.)}\pm 0.21{\rm (syst.)})\times10^{-5}$.
The statistical significance of the combined measurement is estimated by performing a simultaneous fit to the two disjoint datasets, using the same parameterizations as before, and constraining the signal normalization so that the ratio of the signal yield divided by the signal efficiency and the secondary branching fractions is the same in the two datasets.
The statistical significance of the combined measurement is $5.8\sigma$; this is reduced to $5.7\sigma$ when considering yield-related systematic uncertainties by convolving the likelihood function with a Gaussian whose width equals the systematic uncertainty.
This measurement represents the first observation of the hadronic transition $\Upsilon(4S)\to\eta'\Upsilon(1S)$.

We also determine the ratios of branching fractions:
\begin{equation}
R_{\eta'/h}=\frac{{\cal B}(\Upsilon(4S)\to\eta'\Upsilon(1S))}{{\cal B}(\Upsilon(4S)\to h\Upsilon(1S))},
\end{equation}
where the decay is mediated by a hadronic state $h=\eta$ or $\pi^+\pi^-$. For ${\cal B}(\Upsilon(4S)\to h\Upsilon(1S))$, we use the values obtained in Ref.~\cite{ref:Belle4S}, which analyzes  the same data sample considered in this paper. Several systematic uncertainties cancel, being common to the numerator and denominator.
The results from the two $\eta'$ decay modes are combined in a weighted average, as for the branching fraction measurement, and are $R_{\eta'/\eta}=0.20\pm0.06$ and $R_{\eta'/\pi^+\pi^-}=0.42\pm0.11$. The former ratio, in particular, is in agreement with the expected value in the case of an admixture of a state containing light quarks in addition to the $b\bar b$ pair in the $\Upsilon(4S)$ in bottomonium hadronic transitions~\cite{ref:Voloshin}.

The past few years have seen a large amount of activity by both experiment and theory to study precisely the unexpected nature of $\eta$ transitions between bottomonium states. Following this path, the described measurement, being the first observation of an $\eta'$ transition between bottomonia, sheds new light in the comprehension of hadronic transitions.

%***** Acknowledgments *****
We thank the KEKB group for excellent operation of the
accelerator; the KEK cryogenics group for efficient solenoid
operations; and the KEK computer group, the NII, and 
PNNL/EMSL for valuable computing and SINET5 network support.  
We acknowledge support from MEXT, JSPS and Nagoya's TLPRC (Japan);
ARC (Australia); FWF (Austria); NSFC and CCEPP (China); 
MSMT (Czechia); CZF, DFG, EXC153, and VS (Germany);
DST (India); INFN (Italy); 
MOE, MSIP, NRF, RSRI, FLRFAS project and GSDC of KISTI (Korea);
MNiSW and NCN (Poland); MES under contract 14.W03.31.0026 (Russia); ARRS (Slovenia);
IKERBASQUE and MINECO (Spain); 
SNSF (Switzerland); MOE and MOST (Taiwan); and DOE and NSF (USA).

\end{document}